\documentclass[conference]{IEEEtran}
\IEEEoverridecommandlockouts

\usepackage{cite}
\usepackage{amsmath,amssymb,amsfonts}
\usepackage{algorithmic}
\usepackage{multirow}
\usepackage{graphicx}
\usepackage{textcomp}
\usepackage{xcolor}
\def\BibTeX{{\rm B\kern-.05em{\sc i\kern-.025em b}\kern-.08em
    T\kern-.1667em\lower.7ex\hbox{E}\kern-.125emX}}
\begin{document}

\title{Pureformer-VC: Non-parallel Voice Conversion with Pure Stylized Transformer Blocks and Triplet Discriminative Training
}

\author{\IEEEauthorblockN{1\textsuperscript{st} Wenhan Yao}
\IEEEauthorblockA{\textit{School of Computer Science } \\
\textit{Xiangtan University}\\
Xiangtan, China \\
wenhanyao@smail.xtu.edu.cn}
\and
\IEEEauthorblockN{2\textsuperscript{nd} Fen Xiao}
\IEEEauthorblockA{\textit{School of Computer Science } \\
\textit{Xiangtan University}\\
Xiangtan, China \\
xiaof@xtu.edu.cn}
\and
\IEEEauthorblockN{3\textsuperscript{rd} Xiarun Chen}
\IEEEauthorblockA{\textit{School of Software Microelectronics} \\
\textit{Peking University}\\
Beijing, China \\
xiar\_c@pku.edu.cn}
\and
\IEEEauthorblockN{4\textsuperscript{th} Jia Liu}
\IEEEauthorblockA{\textit{School of Software Microelectronics} \\
\textit{Peking University }\\
Beijing, China   \\
2201120008@stu.edu.cn
}
\and
\IEEEauthorblockN{5\textsuperscript{th} YongQiang He}
\IEEEauthorblockA{\textit{School of Software Microelectronics} \\
\textit{Peking University}\\
Beijing, China \\
heyongqiang@stu.pku.edu.cn}
\and
\IEEEauthorblockN{6\textsuperscript{th} Weiping Wen}
\IEEEauthorblockA{\textit{School of Software Microelectronics} \\
\textit{Peking University}\\
Beijing, China \\
weipingwen@pku.edu.cn}
}

\maketitle

\begin{abstract}
As a foundational technology for intelligent human-computer interaction, voice conversion (VC) seeks to transform speech from any source timbre into any target timbre. Traditional voice conversion methods based on Generative Adversarial Networks (GANs) encounter significant challenges in precisely encoding diverse speech elements and effectively synthesising these elements into natural-sounding converted speech. To overcome these limitations, we introduce Pureformer-VC, an encoder-decoder framework that utilizes Conformer blocks to build a disentangled encoder and employs Zipformer blocks to create a style transfer decoder. We adopt a variational decoupled training approach to isolate speech components using a Variational Autoencoder (VAE), complemented by triplet discriminative training to enhance the speaker's discriminative capabilities. Furthermore, we incorporate the Attention Style Transfer Mechanism (ASTM) with Zipformer's shared weights to improve the style transfer performance in the decoder. We conducted experiments on two multi-speaker datasets. The experimental results demonstrate that the proposed model achieves comparable subjective evaluation scores while significantly enhancing objective metrics compared to existing approaches in many-to-many and many-to-one VC scenarios. 

\end{abstract}


\section{Introduction}

Voice conversion (VC) seeks to transform the speaker's timbre in speech to match that of a target speaker while preserving the original content. The task is typically text-independent. Traditional parallel VC research has primarily focused on feature matching methods \cite{toda2007voice,helander2010voice,wu2020one,erro2009voice,aihara2014exemplar}, with the corpus consisting of paired utterances that have identical linguistic content. Consequently, these methods struggle to address the challenge of converting between a wide range of different timbres and poor speech quality. Recently, researchers have increasingly concentrated on non-parallel VC trained on multi-speaker datasets featuring randomly spoken utterances. Non-parallel VC encompasses many-to-many and one-to-many VC tasks, designed to generate diverse timbres or a specific target timbre from various source timbres. In this scenario, the target timbre only minimally participates or does not engage in the VC model training.

Drawing inspiration from the concept of image style transfer in computer vision, generative adversarial networks (GANs) have surfaced as a formidable tool for achieving non-parallel voice conversion (VC). Several GAN-based VC methods have been proposed\cite{kaneko2020cyclegan,kaneko2021maskcyclegan,kameoka2018stargan,kaneko2019stargan,li2021starganv2,chen2021towards}, which do not require explicit parallel target utterances for training. Instead, a discriminator assesses whether a GAN-based VC model produces speech that embodies the target voice characteristics. Consequently, GAN-based VC models learn to convert voice across trained timbres, resulting in limited timbre targets. Within these methods, speaker encoders and style transfer functions play an essential role. They assist the generator in understanding transformation relationships between various speaker domains, which is achieved through the integration of style transfer modules into the generators, such as Adaptive Instance Normalization (AdaIN) \cite{huang2017arbitrary} and Weight Adaptive Instance Normalization (WadaIN) \cite{karras2020analyzing}.

However, training GAN models remains challenging due to issues with convergence and sensitivity to dataset imbalances. In recent years, flow-based VC methods (such as SoftVC-VITS \cite{van2022comparison} and YourTTS \cite{casanova2022yourtts}), KNN-based VC approaches (like KNN-VC \cite{baas2023voice} and its derivative project RVC\footnote{https://github.com/RVC-Project/Retrieval-based-Voice-Conversion-WebUI}), and Generative Large Language-based VC (GLL-VC), including GPT-SoVITS\footnote{https://github.com/RVC-Boss/GPT-SoVITS}, have significantly improved audio quality and training stability. Flow-matching VC and some GLL-VC techniques utilize invertible flow architectures to transform the timbre in either the frequency domain of speech or the discrete speech unit. The KNN-VC method substitutes the source content units with the nearest content units from the target timbre match set. These high-quality VC methods require deep speech unit learning with a multi-speaker corpus pre-training, which can be time-consuming.


Considering that speech can be broken down into multiple components 
\cite{qian2020unsupervised} (e.g., timbre, pitch, content, and rhythm), disentanglement-based VC appears to be a promising approach. This framework enables neural networks to develop distinct representations of each speech component using several encoders and a decoder 
\cite{qian2020unsupervised, chan2022speechsplit2}. During training, each encoder analyzes the corresponding spectrogram to create independent representations of the speech components. The decoder then combines these components to reconstruct the original speech. However, current methods—such as the forced decomposition in SpeechSplit 
\cite{qian2020unsupervised, chan2022speechsplit2}, INVC 
\cite{chou2019one}, and the information bottleneck strategy in AutoVC 
\cite{qian2019autovc}—do not ensure perfect disentanglement or high-quality reconstruction.

\textit{So, how should an efficient encoder-decoder framework be constructed for VC tasks?}

We propose that an effective and practical disentangled voice conversion framework, based on an encoder-decoder architecture, must be fundamentally grounded in three essential principles: (1) encoders and decoders with distinct roles, (2) an optimization objective that enhances representational discriminability, and (3) an efficient style transfer module within the decoder to merge speech components and enable precise speech reconstruction. In recent years, the most effective architectures in the field of speech model backbones have been several enhanced transformer-based networks, such as Conformer\cite{gulati2020conformer}, Paraformer\cite{gao2022paraformer}, and Zipformer\cite{yao2023zipformer}. These models have shown exceptional sequence modeling capabilities and have achieved significant success in applications such as automatic speech recognition\cite{an2024paraformer}, speaker verification\cite{zhang2022mfa,liu2022mfa}, and speech enhancement\cite{kim2021se,abdulatif2024cmgan}. Consequently, we believe that constructing a VC framework utilizing Transformer-based networks is feasible.

Building on the previous discussion, we present \textbf{Pureformer-VC}\footnote{https://github.com/ywh-my/PureformerVC} as a comprehensive solution for a practical VC framework with three technical approaches. For the first approach, we design a specialized content encoder that combines Conformer blocks with IN operations. This structure enhances the model's ability to represent linguistic information through normalized distributions while filtering out speaker characteristics. For the decoder, we utilize Zipformer blocks, which have shown exceptional performance in speech acoustic modeling tasks, to ensure high-quality synthesis output. For the second approach, we integrate the Attention Style Transfer Mechanism (ASTM)
ootnote{wu2021styleformer} within the Zipformer blocks, effectively incorporating speaker information into generated speech. The speaker encoder is likewise constructed using Conformer blocks but omits IN to prevent the potential filtering of speaker information. For the final approach, we introduce a triplet loss \cite{hermans2017defense} alongside the reconstruction loss, allowing the model to learn and maintain distinct distances between utterances of different timbres. In summary, this paper's main contributions are as follows.
\begin{itemize}
  \item We proposed a one-shot, many-to-many VC framework called Pureformer-VC. The key modules are constructed using advanced speech encoding blocks, which assist in preserving the reconstruction quality.

  \item To enhance style transfer, the shared weights in Zipformer are implemented using the ASTM in Styleformer.

  \item We conducted one-shot and many-to-many voice conversion experiments on the VCTK and AISHELL-3 datasets. The evaluation results indicate that our proposed method achieves comparable or even superior results in various voice conversion scenarios compared to existing methods.
\end{itemize}

\section{Related Work}
\subsection{Voice Conversion}

Voice conversion (VC) model training can be broadly categorized into parallel and non-parallel approaches. Early parallel methods relied on utterances with identical content but varying timbres to map features. Techniques such as Gaussian Mixture Models (GMM-VC)\cite{toda2007voice}, Directional Kernel Partial Least Squares (DKPLS)\cite{helander2010voice}, Vector Quantization-based VC (VQ-VC)\cite{wu2020one}, frequency warping\cite{erro2009voice}, and Non-Negative Matrix Factorization (NMF)\cite{aihara2014exemplar} were frequently employed. However, these approaches often yielded overly smooth outputs and demonstrated weak generative performance. Recent technological advancements have triggered a paradigm shift toward non-parallel voice conversion, which can be systematically classified into two primary research directions:

\noindent\textbf{Domain Transfer.} Domain transfer-based VC models, such as the series models of StarGAN-VC and CycleGAN-VC, treat each timbre as a domain and employ cyclic adversarial training to transform features across domains. This method facilitates the generation of more realistic and diverse speech outputs. A framework based on generative adversarial networks (GANs) has proven effective for non-parallel conversion.

\noindent\textbf{Information Disentanglement.} Information disentanglement seeks to break down speech into distinct components through an encoder, including content, timbre, pitch, and rhythm, thereby allowing flexible recombination by a decoder. Representative models, such as INVC, SpeechSplit, and MAIN-VC \cite{li2024main}, utilize techniques like IN and mutual information estimation to achieve effective disentanglement and reconstruction.

\noindent\textbf{Generative VC.} We consider recent VC methods based on generative theory to be generative VC approaches, which include flow-based VC \cite{van2022comparison,casanova2022yourtts}, KNN-based VC \cite{baas2023voice,shao2025knn}, and GLL-VC (including GPT-SoVITS, etc.). In the voice conversion stage, flow-based models can invert the semantics of speech into textual information and then convert it back into speech with a new timbre. The KNN-based VC replaces the source speech's deep speech unit vectors with the given target speech. The GLL-VC models pretrain the discrete speech semantic representations using regression training and predict different speech timbres based on reference speech prompts.

This evolution from parallel to non-parallel methods emphasizes the shift toward models with enhanced flexibility, robustness, and generative capabilities in VC tasks. Formally, most non-parallel VC models can be represented as $y=G(x_{s},x_{y})$, where $x_{s}$ represents the speech providing the content and $x_{y}$ represents the speech supplying the speaker's voice characteristics.

\subsection{Style Transfer Learning in VC}

Style transfer learning teaches VC models to merge various speech representations. Accordingly, the style transfer function accepts both source speaker-independent and target speaker-dependent representations. Chou et al. \cite{chou2019one} were the first to discover that IN can filter out speaker information while retaining the source content from original utterances in INVC. Subsequently, the IN function found widespread application in GANs-based VC. \cite{kaneko2019stargan,li2021starganv2}. Furthermore, the WadaIN method implements affine operations on the convolutional kernel in Convolutional Neural Networks (CNNs), modifying the style of source data in WadaIN-VC \cite{chen2021towards} by convolving the source data. However, these models depend on CNN architectures and exhibit limited reconstruction capabilities. To leverage the self-attention mechanism in Transformers, the Attention-AdaIN-VC \cite{ke2022new} incorporated the styleformer block within the CNN blocks, achieving improved VC performance. In styleformer \cite{wu2021styleformer}, self-attention weights are stylized using speaker representations, successfully training an image style transfer model. In the styleformer's blocks, ASTM is commonly used to integrate individual embedding with self-attention layers. We typically incorporate ASTM into our decoder as a style transfer module.

\begin{figure*}[ht]
    \centering
    \includegraphics[scale=0.8]{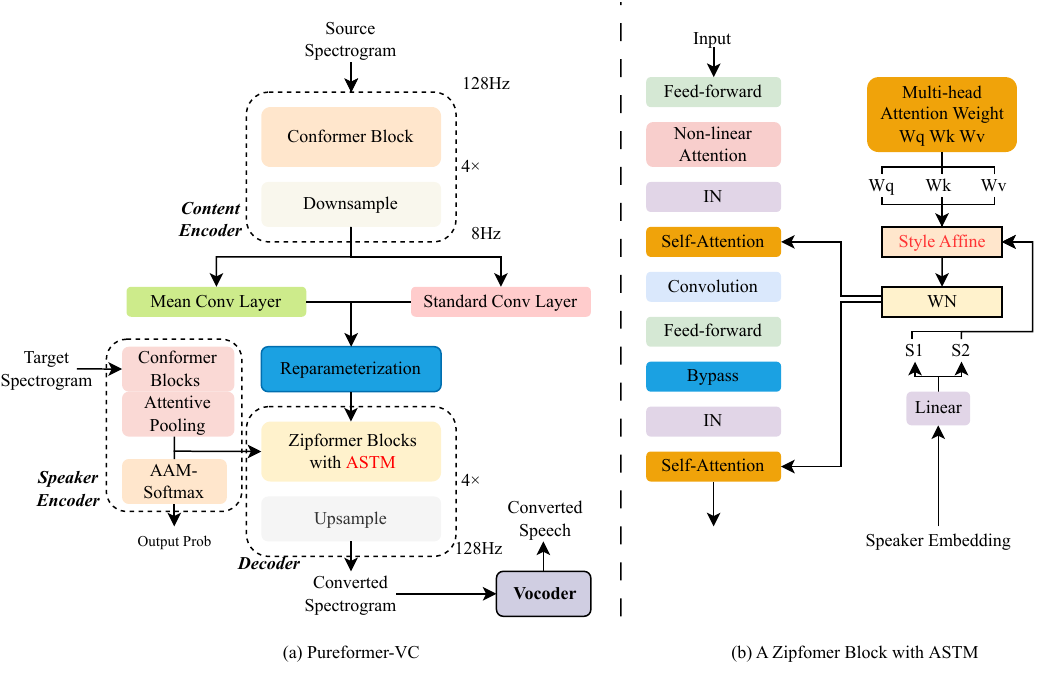}
    \caption{The architecture of Pureformer-VC.}
    \label{fig:1}
\end{figure*}

\section{Methodology}

\subsection{Overall Architecture}
The overall architecture of Pureformer-VC is depicted in Figure \ref{fig:1}(a). Pureformer-VC consists of a content encoder, a decoder, a speaker encoder, and a vocoder. We utilized a pre-trained Hifi-GAN generator \cite{kong2020hifi} as the vocoder, which remains frozen during the training stage. 

We set the input feature mel-spectrogram as $x \in X^{[L, D]}$, where L denotes the frame number and D represents the number of mel filters. The content encoder $E_{c}$ extracts the posterior variances of the content representation $r_{m}, r_{s}=E_{c}(x)$. The speaker encoder $E_{s}$ generates speaker embeddings as the timbre representation $s = E_{s}(x)$ from mel-spectrograms. The decoder $E_{d}$ takes the reparameterization variance and outputs the converted spectrogram $x_{dec}$, incorporating the style embedding for transfer. It is important to note that \textit{e} is a random variable that follows the standard normal distribution $N(0,1)$.

\subsection{Content Encoder with VAE Training}
The content encoder parameterizes and approximates the variational distribution of $q_{\phi}(z|x)$. Each Conformer block was constructed \textbf{with an IN function} and an AveragePooling1D layer following the convolution module to reduce the time dimension by half. There are four continuous blocks as shown in Figure \ref{fig:1} (a); thus, the output spectrogram's frame length is decreased by 16. Finally, the content encoder outputs the reparameterization of the content representation by two convolution layers:

\begin{align}
    r_m &= Mean\_Conv(h_{enc}) \\
    r_{s} &= Std\_Conv(h_{enc}) \\
    r_{c} &= r_{m} + \textit{e}*r_{s}
\end{align}

The $Mean\_Conv$ and $ Std\_Conv$ denote the two convolution layers after the output $h_{enc}$ of Conformer blocks. Thus, the variable $r_{c}$ follows a normal distribution with mean $ r_{m}$ and variance $r_{s}$. The reparameterization operation ensures that the speech latent variables input into the decoder remain typically distributed while also preserving the gradient propagation in the model. It turns out to optimize the Evidence Lower Bound (ELBO)\cite{zhu2020batch} of $log (p(x))$:

\begin{equation}
    L_{elbo} = E[logP_{\theta}(x|z)] - KL(q_{\phi}(z|x) || p(z))
\end{equation}
where $\phi$ denotes the encoder network and $\theta$ represents the decoder. The first term above is the
reconstruction loss, while the second is the KullbackLeibler divergence between the approximate posterior and the prior. Thus, the VAE training loss can be summarized as:

\begin{align}
 L_{vae}(x,x_{dec}) = &E\left[ |x-x_{dec}| \right] + \\
 0.5 \cdot &E[r_{c} + r_{m}^2 - log(r_{m}^2) - 1 ] 
\end{align}

The $r_{c}$ is derived from the input mel-spectrogram $x$.

\subsection{ASTM in Decoder}
ASTM aids the decoder in learning the timbral characteristics of the target speech. We built the decoder using 4 Zipformer blocks with ASTM, as illustrated in Figure \ref{fig:1}(b). The decoder merges the content and timbre representations of the speech. 

To generate speech with diverse voice styles, we apply the ASTM to the weights in the self-attention mechanism of Zipformer blocks. During the model initialization phase, the ASTM initially sets some attention weights $w_{q}, w_{k}, w_{v}, w_{u}$. These weights are infused with the style characteristics of the split speaker embedding vector $s_{1}, s_{2} = split(E_{s}(x))$ as follows:

\begin{align}
    w_{q} = w_{q} \cdot s_{1} + s_{2}, w_{k} = w_{k} \cdot s_{1} +s_{2}\\
    w_{v} = w_{v} \cdot s_{1} + s_{2}, w_{u} = w_{u} \cdot s_{1} +s_{2}
\end{align}

 Then, we apply weight normalization (WN) 
\cite{salimans2016weight} to the weights to achieve improved convergence performance. WN takes the weight $w$ and normalizes it at the output dimension $i,j$ as follows:

 \begin{align}
    w_{ij}' = w_{ij} \cdot \frac{1}{\sqrt{w_{ij}^2}}
\end{align}
 
 Using the WN operation, we scale the output of each weight $w \in \{ w_{q},w_{k},w_{v},w_{u} \}$ back to a unit standard deviation. The WN aids the model in accelerating training convergence following the attention calculation with stylized weights. Consequently, in the self-attention layers, the stylized attention is as follows:

  \begin{align}
    x' &= norm(s_{1} \cdot x + s_{2}) \\
    out &= \frac{wn(w_{q})x' \cdot (wn(w_{k})x')^{t}}{\sqrt{d}} \cdot wn(w_{k})x' + wn(w_{u})x'
\end{align}

The $norm$ represents a non-parameterized layer normalization function. Additionally, $d$ indicates the output dimension for each weight, and $wn$ refers to the WN. The stylized attention output is connected residually.

\subsection{Speaker Encoder with AAM-Softmax Loss}
The speaker encoder is designed to extract timbre representations from mel-spectrograms, allowing the model to capture speaker-specific features that are critical for VC. For its backbone, we use a structure made up of several Conformer blocks from MFA-Conformer\cite{zhang2022mfa}. Importantly, these Conformer blocks are configured \textbf{without the IN functions} to ensure the preservation of speaker-related information.

To further enhance the quality of the extracted embeddings, we incorporate the Additive Angular Margin Softmax (AAM-softmax) layer. This parameterized loss function optimizes the learning of compact and well-separated clusters in the embedding space for different speakers. By introducing a fixed angular margin between classes, the AAM-softmax layer encourages the encoder to produce discriminative and robust embeddings, thus improving the overall performance of speaker representation in the VC process.

\subsection{Triplet loss and Data Sample Strategy}
Considering the previous disentanglement-based VC models, both the source and target mel-spectrograms were identical during the training stage but differed during the inference stage. This discrepancy between training and inference diminishes the model's generalization capability and results in poor speech quality when the target spectrograms lack speaker information. To address this issue, we utilize the triplet loss, as shown in Figure \ref{fig:2}, which is an unsupervised learning technique featuring discriminative training that allows the speaker encoder to discern the differences in timbre among various voices. The triplet loss training necessitates a unique data sampling strategy. 

During the training stage, we sample three utterance segments of equal length from the dataset: an anchor sample $x_{anc}$, a positive sample $x_{pos}$, and a negative sample $x_{neg}$. As illustrated in Figure \ref{fig:2}, the anchor sample and the positive sample share the same timbre, while the negative sample is from a different speaker than the anchor. Therefore, let $nm$ represent the L2 normalization. We can use the speaker encoder outputs of the three samples to calculate a triplet loss:

\begin{align}
    e_{anc},e_{pos},e_{neg} = E_{s}(x_{anc}),E_{s}(x_{pos}),E_{s}(x_{neg})
\end{align}

\begin{align}
     L_{tri} = E[nm(e_{anc}) * nm(e_{pos})^t] - \\\notag   E[nm(e_{anc}) * nm(e_{neg})^t]  + \delta
\end{align}
The $\delta$ is a hyper-parameter to control the speaker similarity. Denoting the VC model as $G_{vc}$, the total model's output can be described as follows:

\begin{align}
    y_{1} = G_{vc}(x_{anc},x_{neg})\\
    y_{2} = G_{vc}(x_{anc},x_{pos})
\end{align}

\subsection{Training Objective}
\begin{figure}[t]
    \centering
    \includegraphics[scale=0.35]{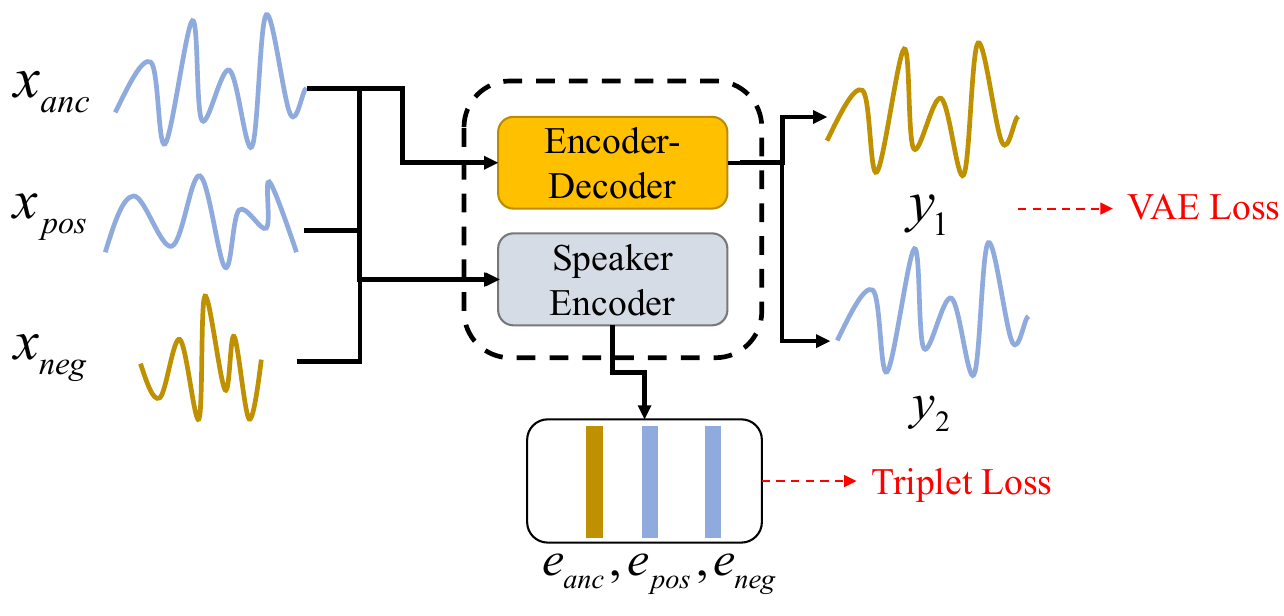}
    \caption{The illustration of training objective.}
    \label{fig:2}
\end{figure}

The training objective of the Pureformer-VC model includes VAE loss, AAM-softmax loss, and triplet loss. The total VAE loss is based on two outputs $y_{1},y_{2}$ as shown in Figure \ref{fig:2} and can be denoted as:
\begin{align}
     L_{t-vae} = \lambda_{1}(L_{vae}(x_{anc},y_{1}) + \lambda_{2}L_{vae}(x_{anc},y_{2}))
\end{align}

The AAM-softmax loss can be computed using the three true labels of samples $C$ and the predictions of the speaker encoder:
\begin{equation}
    L_{t-aam} = \sum_{c_{i}\in C} L_{aam}(c_{i},x_{i})
\end{equation}

Finally, the triplet loss helps the speaker encoder to distinguish the embeddings. The total training objective is as follows:
\begin{equation}
    L_{total} = L_{t-vae} + \lambda_{3}L_{t-aam} + \lambda_{4}L_{tri} 
\end{equation}

\subsection{Vocoder}
The vocoder has the same structure as the HiFi-GAN generator. In our study, it was pre-trained on the same dataset used for the VC training.

\section{Experiments and Results}
\subsection{Experimental Setup}
\noindent\textbf{Datasets and Feature Setup.} To evaluate the effectiveness of Pureformer-VC, we conducted a comparative experiment and an ablation study on VCTK\cite{Veaux2020} and AISHELL-3\cite{shi2020aishell} datasets. The VCTK corpus includes 109 English speakers, each reading about 400 utterances. The AISHELL-3 corpus contains roughly 85 hours of emotion-neutral recordings spoken by 218 native Chinese Mandarin speakers and 88035 utterances. 

The mel-spectrogram extraction process must align with the pre-trained vocoder and adhere to the algorithm outlined in the HiFi-GAN framework\cite{kong2020hifi}. The signal hyperparameters are defined as follows: 80 Mel frequency filters, 1024 FFT bins, a window length of 1024, a hop length of 256, a sampling rate of 22,050 Hz, and a maximum frequency of 8 kHz. The final input features are log-mel spectrograms, obtained by applying the logarithm to the extracted mel-spectrograms.

Considering batch sampling during the training stage, we randomly selected an utterance from one speaker. We then sampled two utterances from another speaker to create a training sample: $\{ x_{anc},x_{pos},x_{neg} \}$. Five speakers are randomly chosen as unseen speakers for each corpus.



\begin{table*}[t]
\centering
\caption{
Comparison of baseline and proposed methods for many-to-many and one-shot VC on the VCTK dataset (with a $95\%$ confidence interval).}
\label{table:1}
\begin{tabular}{ccccccc}
\hline
\multirow{2}{*}{VC method}                                            & \multicolumn{3}{c}{Many-to-many} & \multicolumn{3}{c}{One-shot} \\ \cline{2-7} 
 & MCD$\downarrow$         & MOS $\uparrow$      & VSS$\uparrow$      & MCD$\downarrow$      & MOS$\uparrow$    & VSS$\uparrow$  \\ \hline
AdaIN-VC\cite{chou2019one} &7.64 ± 0.24            &3.11 ± 0.13         &2.82 ± 0.19         &7.38 ± 0.14           &3.04 ± 0.21    &2.45 ± 0.16      \\
AutoVC\cite{qian2019autovc} &6.68 ± 0.21           &2.76 ± 0.16          &2.45 ± 0.23          &8.08 ± 0.15          &2.68 ± 0.17         &2.39 ± 0.14        \\
VQMIVC\cite{wang2021vqmivc} &6.24 ± 0.13           &3.20 ± 0.14          &3.32 ± 0.12          &5.59 ± 0.10          &3.16 ± 0.18         &3.02 ± 0.18         \\
MAIN-VC\cite{li2024main} &5.28 ± 0.11           &3.44 ± 0.12          &3.25 ± 0.16          &5.42 ± 0.13          &3.24 ± 0.18         &3.29 ± 0.11      \\   

RVC$^{1}$   & 4.75 ± 0.09         & 3.70 ± 0.07          &  3.72 ± 0.10       & 4.70 ±  0.12         & 3.41 ± 0.08        & 3.47 ± 0.09  \\
GPT-SoVITS$^{2}$ & 4.70 ± 0.07          & 3.70 ± 0.08         & 3.68 ± 0.10        &  4.68 ± 0.12       & 3.43 ± 0.09       & 3.50 ± 0.10  
\\ \hline
\begin{tabular}[c]{@{}l@{}}Pureformer-VC(w/o AAM-softmax)\end{tabular} &5.25 ± 0.11           &3.42 ± 0.12          &3.05 ± 0.16          &5.21 ± 0.12          &3.20 ± 0.11         &3.10 ± 0.11        \\

Pureformer-VC(w/o triplet) &5.06 ± 0.09           &3.25 ± 0.12          &3.55 ± 0.14          &4.92 ± 0.11          &3.32 ± 0.15         &3.21 ± 0.10         \\

Pureformer-VC &4.95 ± 0.12           &3.64 ± 0.13          &3.56 ± 0.13          &4.89 ± 0.10          &3.36 ± 0.15         &3.35 ± 0.12         \\

\hline

\end{tabular}
\end{table*}



\begin{table*}[t]
\centering
\caption{
Comparison of baseline and proposed methods for many-to-many and one-shot VC on the AISHELL-3 dataset (with a $95\%$ confidence interval).}
\label{table:2}
\begin{tabular}{ccccccc}
\hline
\multirow{2}{*}{VC method}                                            & \multicolumn{3}{c}{Many-to-many} & \multicolumn{3}{c}{One-shot} \\ \cline{2-7} 
 & MCD$\downarrow$       & MOS$\uparrow$      & VSS$\uparrow$      & MCD$\downarrow$      & MOS$\uparrow$    & VSS$\uparrow$  \\ \hline
AdaIN-VC\cite{chou2019one} &7.29 ± 0.34            & 3.23 ± 0.21      & 2.76 ± 0.15       &6.95 ± 0.20     &3.13 ± 0.17    & 2.34 ± 0.11   \\
AutoVC\cite{qian2019autovc} &6.26 ± 0.25        &2.85 ± 0.15          &2.40 ± 0.19          &6.64 ± 0.23         &3.21 ± 0.20        &2.36 ± 0.15       \\
VQMIVC\cite{wang2021vqmivc}  &6.07 ± 0.16          &3.18 ±  0.17        & 3.34 ± 0.24         &5.35 ± 0.16          &3.02 ± 0.24        & 3.13 ± 0.11         \\
MAIN-VC\cite{li2024main}  &5.06 ± 0.14           &3.56 ± 0.16          &3.29 ± 0.09          &5.26 ± 0.11          &3.35 ± 0.16        &3.32 ± 0.17   \\   

RVC$^{1}$  & 4.36 ± 0.12        &  3.74 ± 0.11        &  3.65 ± 0.12       &  4.34 ± 0.08        &   3.65 ± 0.11      & 3.60 ± 0.13  \\
GPT-SoVITS$^{2}$  & 4.28 ± 0.09          & 3.75 ± 0.14          & 3.69 ± 0.15        & 4.30 ± 0.13           & 3.68 ± 0.12        & 3.63 ± 0.12

\\ \hline
\begin{tabular}[c]{@{}l@{}}Pureformer-VC(w/o AAM-softmax)\end{tabular} & 4.97 ± 0.16          &3.58 ± 0.14         &3.27 ± 0.12         &5.05 ± 0.13          &   3.37 ± 0.12     & 3.28 ± 0.16       \\

Pureformer-VC(w/o triplet) &4.66 ± 0.20           &3.65 ± 0.15          &3.32 ± 0.17          &4.94 ± 0.22         & 3.45 ± 0.15       & 3.39 ± 0.24      \\

Pureformer-VC &4.50 ± 0.15          &3.69 ± 0.20        &3.56 ± 0.19          &4.64 ± 0.14          & 3.59 ± 0.16     &3.50 ± 0.18        \\

\hline

\end{tabular}
\end{table*}

\noindent\textbf{Training Setup.} In the training stage, the batch size is 16. The learning rate is constant at $2 \times 10^{-4}$. The Pureformer-VC is trained by Adam optimizer\cite{kingma2014adam} with $ \beta_{1} = 0.9, \beta_{2} = 0.99, \epsilon = 1 \times 10^{-6}$. The $\lambda_{1}$ is set to 10 and $\lambda_{2}$ ranges from $1 \times 10-4 $ to 1. Both the $\lambda_{3},\lambda_{4}$ are set to 1. The $\delta$ is 0.3.

\noindent\textbf{Baseline Setup.} We compared Pureformer-VC with recent VC frameworks, such as AdaIN-VC\cite{chou2019one}, AutoVC\cite{qian2019autovc}, VQMIVC\cite{wang2021vqmivc}, MAIN-VC\cite{li2024main}, RVC$^{1}$, and GPT-SoVITS$^{2}$. The experiments are conducted in a many-to-many, one-shot (any-to-any) setup. We further evaluate the performance of the proposed model in cross-lingual VC, where the source utterance's language differs from that of the target language.

\subsection{Metrics and Evaluation}
We assess the naturalness and intelligibility of the generated speech using subjective metrics, such as the Mean Opinion Score (MOS). Additionally, we utilize objective metrics to evaluate timbre similarity with the target speech, including the Voice Similarity Score (VSS) and Mel-Cepstral Distortion (MCD). Higher scores indicate greater effectiveness of the voice conversion (VC) system. 

\noindent\textbf{Mean Opinion Score (MOS)}. The Mean Opinion Score (MOS) is a widely used metric for assessing the subjective quality of speech or audio. It is based on ratings from human listeners, who are asked to evaluate the quality of speech samples using a scale that typically ranges from 1 to 5. A higher MOS signifies better reconstruction quality.

\noindent\textbf{Voice similarity score (VSS)}. The Voice Similarity Score (VSS) is an objective metric that quantifies the degree of resemblance between generated speech and authentic or target speech in terms of timbre, tone, and voice quality. VSS is calculated based on embedding similarity derived from a pre-trained speaker verification model (Resemblyzer)
\cite{desplanques2020ecapa}. Higher scores represent greater similarity, indicating improved voice conversion (VC) performance.  

\noindent\textbf{Mel-cepstral distortion (MCD)}. The MCD is an objective quantitative measure that evaluates the Mel-cepstral divergence between the source and generated utterances. The lower the MCD, the better the reconstruction effect.

During the testing phase, both many-to-many and one-shot (any-to-any) VC are conducted using non-parallel data. For evaluation, 10 source/target speech pairs are fed into each VC model under two scenarios. After this, 5 participants are invited to rate the speech samples.

\subsection{Experimental Results}
Table \ref{table:1} and \ref{table:2} present a comparative analysis of the Pureformer-VC against baseline methods in both many-to-many and one-shot VC settings across both datasets. 

\noindent\textbf{MCD and MOS metrics.} The MOS and MCD can evaluate the effectiveness of speech reconstruction, collectively referred to as reconstruction metrics. Since the original sampling rate of the AISHELL-3 dataset is higher than that of the VCTK dataset, the quality of speech reconstruction is superior, resulting in better reconstruction metrics for the AISHELL-3 dataset. The Pureformer-VC outperforms the encoder-decoder baseline models in both many-to-many and one-shot experimental configurations with respect to reconstruction metrics. However, when compared to state-of-the-art methods like RVC and GPT-SoVITS, the Pureformer-VC still exhibits a slight performance gap. These results showcase the effectiveness of the pure transformer architecture.

\noindent\textbf{VSS metric.} The VSS evaluation measures the similarity between the generated target timbre and the actual target timbre using a resemblyzer. Compared to the four classic encoder-decoder-based VC baselines, the Pureformer-VC surpasses them in the VSS metric. However, a slight gap remains in the VSS performance between PVC and the state-of-the-art models RVC and GPT-SoVITS. As shown in the results, we further investigated the impact of removing either the AAMSoftmax loss or the triplet loss from the training objectives to assess the model's ability to represent timbre in the embedded vectors during training. It is evident that without these two losses, the proposed model's VSS stays relatively consistent with that of the baseline model. Therefore, incorporating these losses helps enhance the model's VC expressiveness.

\begin{figure}[ht]
    \centering
    \includegraphics[width=\linewidth]{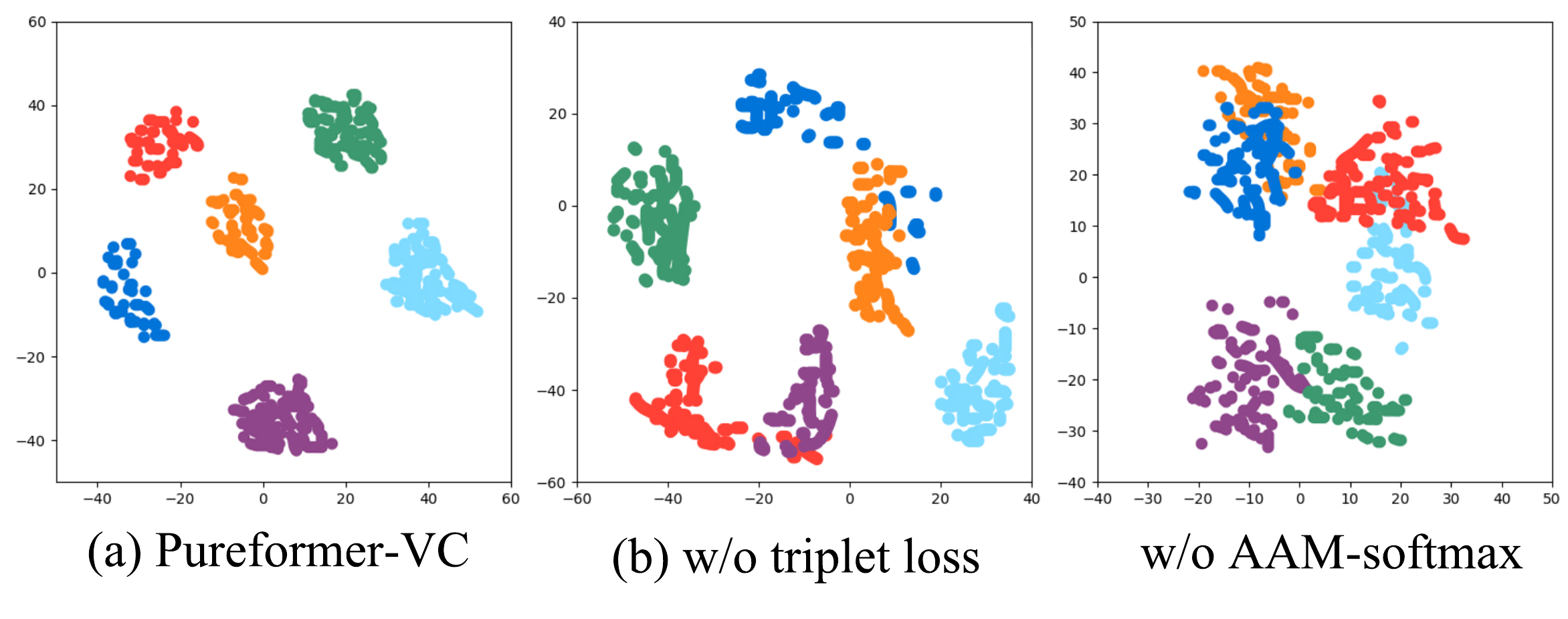}
    \caption{The visualization of speaker representations extracted from 6 unseen
speakers’ utterances.}
    \label{fig:4}
\end{figure}

\subsection{Ablation Study}
We conduct ablation experiments to validate the effects of triplet loss and AAM-softmax loss on disentanglement. We set up the following models: (a) the Pureformer-VC model, (b) the Pureformer-VC model without triplet loss (w/o triplet), and (c) the Pureformer-VC model without AAM-softmax loss (w/o AAM-softmax). We used the resemblyzer to detect synthetic speech and evaluate conversion quality. It assigns detection scores to fake (i.e., the VC model's experimental outputs) and authentic utterances from the target speaker after learning the target's characteristics from ten additional genuine utterances. A higher score signifies a closer resemblance in timbre and superior speech quality. The results are detailed in Table \ref{table:3}. We found that the decision scores of our method are higher than those of the best baseline model, MAIN-VC. Furthermore, the experiments demonstrate that both triplet loss and AAM-Softmax contribute to improving the accuracy of timbre generation.

For a visual evaluation of each model's disentanglement capability, the t-SNE scatter plots of the speaker representations are shown in Figure \ref{fig:4}. The AAM-softmax loss has a significant impact on the clustering of speaker embedding vectors, while the triplet loss helps create more distinct boundaries between categories.


\begin{table}[htbp]
\centering
\caption{ Fake Detection score comparison for ablation study ablation}
\label{table:3}
\begin{tabular}{lcc}
\hline
\multicolumn{1}{c}{Method} & \multicolumn{2}{c}{Detection Score$\uparrow$} \\ \hline
 & VCTK             & AISHELL-3        \\ \hline
Pureformer-VC              & 0.75 ± 0.08      & 0.73 ± 0.08      \\
w/o triplet loss           & 0.65 ± 0.12      & 0.68 ± 0.12      \\
w/o AAM-Softmax            & 0.52 ± 0.12      & 0.56 ± 0.12      \\
MAIN-VC                    & 0.74 ± 0.01      & 0.73 ± 0.01      \\
RVC                    & 0.78 ± 0.01      & 0.75 ± 0.01 \\
GPT-SoVITS                    & 0.80 ± 0.01      & 0.77 ± 0.01
\\ \hline
\end{tabular}
\end{table}

\subsection{Cross-lingual VC}
Pureformer VC is also capable of performing cross-lingual voice conversion (CVC), where the source and target utterances are in different languages. We trained Pureformer-VC on a mixture of the AISHELL-3 and VCTK datasets. Due to varying pronunciation habits, the reconstruction metrics ($MOS=3.15, MCD=5.12$) and VSS ($2.95$) scores for cross-lingual voice conversion are lower than those for monolingual VC experiments. The bilingual timbre experiments suggest that additional latent variables may be necessary to decouple the languages using special encoders for improved conversion performance.

\section{Conclusion}

In this paper, we present a novel approach to voice conversion (VC) by leveraging a pure transformer network designed as a VAE encoder-decoder framework called Pureformer-VC. Within the decoder, we integrate a styleformer module, which enhances the model’s capacity for style transfer. Additionally, we improve the effectiveness of the speaker encoder by incorporating triplet loss and AAMSoftmax loss. These enhancements significantly boost the model's ability to capture and represent the nuances of different speaking voices, resulting in more accurate and robust VC. In conclusion, the Pureformer-VC model, fortified by the strategic application of specialized loss functions and style adaptation mechanisms, represents a significant advancement in the field of VC.

\bibliographystyle{IEEEtran}
\bibliography{template}
\end{document}